\documentclass{aastex6}

\begin{document}


%\title{Hot horizontal branch stars in the metal rich, old open cluster, NGC~6791: binary evolution does work.}
\title{Binarity as the solution to the stellar evolution enigma posed by NGC~6791}

\author{Giovanni Carraro\altaffilmark{1}}
\affil{Dipartimento di Fisica e Astronomia {\it Galileo Galilei} \\
Universit\'a di Padova\\
Vicolo Osservatorio 3, 
I-35122, Padova, Italy}

\and

\author{Omar G. Benvenuto\altaffilmark{2}}
\affil{Instituto de Astrof\'isica de La Plata, CCT-CONICET-UNLP and\\
Facultad de Ciencias Astron\'omicas y Geof\'isicas, Universidad Nacional de La Plata 
Paseo del Bosque S/N, B1900FWA La Plata, Argentina}

\altaffiltext{1}{giovanni.carraro@unipd.it}
\altaffiltext{2}{Member of the Carrera del Investigador Cient\'{\i}fico, Comisi\'on de Investigaciones Cient\'{\i}ficas de la Provincia de Buenos Aires (CIC-PBA), Email: obenvenu@fcaglp.unlp.edu.ar}

%------------------------------------------------------------------
\begin{abstract}

Binary evolution is investigated as the source for the extreme horizontal branch (EHB) stars in the old and metal rich open cluster NGC~6791.   
Employing an updated version of our binary stellar evolution code we demonstrate that EHB stars naturally emerge from the common envelope phase. In sum, the binary model reproduces the observed ($T_{\rm{eff}}$, $\log{g}$) and temporal properties of the EHB over-density tied to NGC 6971, without needing an ad-hoc and anomalous mass-loss prescription.

\end{abstract}

%------------------------------------------------------------------

%------------------------------------------------------------------
\section{Introduction} \label{sec:intro}

NGC~6791 is a metal rich and old Galactic open cluster ([Fe/H] = 0.30-0.40, $\tau \sim7$ Gyr) that exhibits two prominent overdensities on the horizontal branch (HB).  Approximately 45 stars occupy the red clump (RC) region \citep{bu12}, which is reproduced by standard stellar evolution codes modeling metal rich clusters.  However, 12 cluster stars - with membership confirmed only for some of them - are significantly hotter than their RC counterparts, which is in conflict with such classic modeling.  Those stars are designated extreme HB stars (EHB) when associated with star clusters, and hot subdwarf B and potentially O-stars when belonging to the field (sdB/sdO).  Their effective temperature T$_{\rm{eff}}$ and gravity span T$_{\rm{eff}}$ = 25,000$-$45,000 $^{\circ}$K and $\log{g}$ = 4.5$-$6.2 \citet{li94}, respectively. Presumably, these stars are surrounded by a thin hydrogen envelope ($\sim 0.01\ M_{\odot}$).   EHB stars are present in a number of Galactic globular clusters \citep{mbd08}. However, first NGC 6791 HB is much different from any globular cluster HB, as amply discussed in \citep{li94}; second, the combination of mass, age, and metallicity makes NGC 6791 a unique system, with no overlap with Galactic globular clusters.

The mechanism and stellar evolutionary path that gives rise to EHB in NGC 6791 has been an active source of debate, particularly since the discovery of a bimodal HB distribution in NGC 6791 \citep{ku92}.  One proposal involves invoking extreme mass loss that is tied to the cluster's high metallicity \citep{dc96,yo00,ka07}, and whereby a \citet{re75} stellar wind mass-loss parameter as large as $\eta\sim1.2$ is adopted.  Conversely, RC stars exhibit typical masses of 1.03$\pm$0.03 M$_{\odot}$ and lose 0.09$\pm$0.07 M$_{\odot}$ while ascending the red giant branch (RGB) phase \citep{mi12}, which implies a mass-loss compatible with a significantly smaller Reimers parameter of 0.1$ \leq \eta\ \leq $0.3.  Direct observations confirming a sizeable mass-loss rate \citep[e.g., $\sim 10^{-9}$ M$_{\odot}$/yr][]{yo00} remain outstanding.  Astero-seismological studies support a marginal mass-loss rate \citep{mi12}, while direct Spitzer observations did not reveal circum-stellar dust production that would accompany enhanced mass-loss during the RGB phase \citep{va08}. Furthermore, there is a lack of consensus on the details of (fine tuned)  mass-loss required to yield the T$_{\rm{eff}}$ and envelope size of EHB stars. Lastly, current models reproduce nearly all evolutionary phases present in the CMD of NGC~6791 without anomalous mass-loss \citep{cc95,bu12}.

An alternative mechanism reiterated by \citet{li94} and \citet{ca96} is that EHB stars could emerge from type B or C binary systems, whereby their envelope would be largely removed during a common-envelope (CE) phase \citep[see also][]{me76,ha02,br08}. Complex models were not initially readily available to provide a robust evaluation of the hypothesis, and moreover, there was little evidence for binarity among the sample.   Subsequent observations and models in concert suggest that NGC~6791 exhibits a high binary percentage of $\sim$50$\%$ \citep{be08,tw11}, and among the EHB class three systems have been confirmed: B4 \citep{mo03,pa11}, B7 and B8 \citep{mo03,va13}. Indeed, it has been noted that a high-fraction of sdB stars might belong to binary systems \citep[][and references therein]{gr01,ma04}.  

In this study, it is demonstrated that an updated prescription of the \citet{bd03} binary evolutionary code successfully predicts the observed and temporal properties of EHB stars in NGC~6791.
 
%------------------------------------------------------------------
\section{Results} \label{sec:Results}

In the following analysis EHBs are thought to arise from binary evolution, which provides a natural mechanism of depleting the hydrogen-rich outer layer of the star without an ad-hoc or simplified prescription of mass-loss.  Essentially, the mass transfer to the companion is unstable and thus a CE encompasses the stars, subsequently, as the two stellar nuclei approach each other the envelope expands owing to heating and is lost \citep[][see also the discussion in \citealt{ma04}]{pa76}.  

The binary evolution is modelled via an updated version of the \citet{bd03} code, who developed a Henyey-type algorithm to compute stellar evolution in close binary systems, based on a modification of the scheme presented by \citet{kwh67} to solve the set of differential equations of stellar evolution together with the mass transfer rate. This approach was subsequently modified to ameliorate transporting extremely steep chemical profiles outwards (corresponding to stars just prior to undergoing the helium flash).  Convection is treated using the canonical mixing length theory with $\alpha_{mlt}= 2.0$, and semi-convection was introduced following \citet{la85} with $\alpha_{sc}= 0.1$.  

It is known that diffusion slightly affects horizontal branch evolution \citep{2012MNRAS.427.1245R} and it is certainly necessary to account for surface abundances of EHB stars \citep{2007ApJ...670.1178M}. Here, because of the exploratory nature of this paper, diffusion processes were ignored and will be addressed elsewhere.

Donor stars that evolve to EHB conditions should have initial masses marginally larger than that of cluster turn-off, as EHB stars are undergoing core helium burning, which is an evolutionary phase appreciably shorter than core hydrogen burning.  Binary systems consisting of similar mass stars are considered, whereby one star is 1.3~$M_{\odot}$, above the turn-off ($M_{to} \approx 1.15 \; M_{\odot}$), and the companion is slightly below and features a sufficiently lengthy initial orbital period.  The stars are modelled with a metallicity of $Z= 0.04$.   Moreover, EHBs should stem from stars that reached the red giant branch in the recent past.  Consequently, binaries are considered whereby the primaries fill their Roche lobes as they have extended convective envelopes.  Such conditions result in systems that undergo a CE stage in which the primary loses the bulk of its hydrogen rich envelope, while the companion keeps its initial mass and the orbital period falls off appreciably.  EHB stars are the objects that evolve after emerging from the CE phase.

Our main interest is not the CE phase but the emerging objects. So, the CE stage is mimicked assuming a strong mass-loss rate until detachment \citet{it93}. This makes the deep chemical composition profile to remains essentially unaltered, which is expected since CE lasts little time. The binary pair is assumed to undergo the CE phase when the helium core reaches mass values of 0.3480, 0.3694, and 0.4067~$M_{\odot}$. All of them ignite helium well after the CE phase.  Larger helium core masses ignite helium before reaching EHB conditions, and delineate an evolutionary path that is unimportant for the present discussion.

Each model was evolved until reaching a radius of detachment ($R_{d}$) of 7.5 and 1~$R_{\odot}$. For $R_{d}= 7.5\; R_{\odot}$, the total masses corresponding to each helium core at the end of the CE phase were 0.35048, 0.37367, and 0.41344~$M_{\odot}$, whereas for $R_{d}= 1\; R_{\odot}$ the results were slightly smaller, namely 0.34863, 0.37084, and 0.40973~$M_{\odot}$. The differences correspond to the varying thickness of the outermost hydrogen layer. After the CE phase, the stars are evolved at constant mass, and the computations are terminated at an age of $\tau \sim 9$~Gyr, which is an upper limit for the age of NGC~6791. 

\begin{figure} \begin{center}
\includegraphics[scale= 0.60,angle=0]{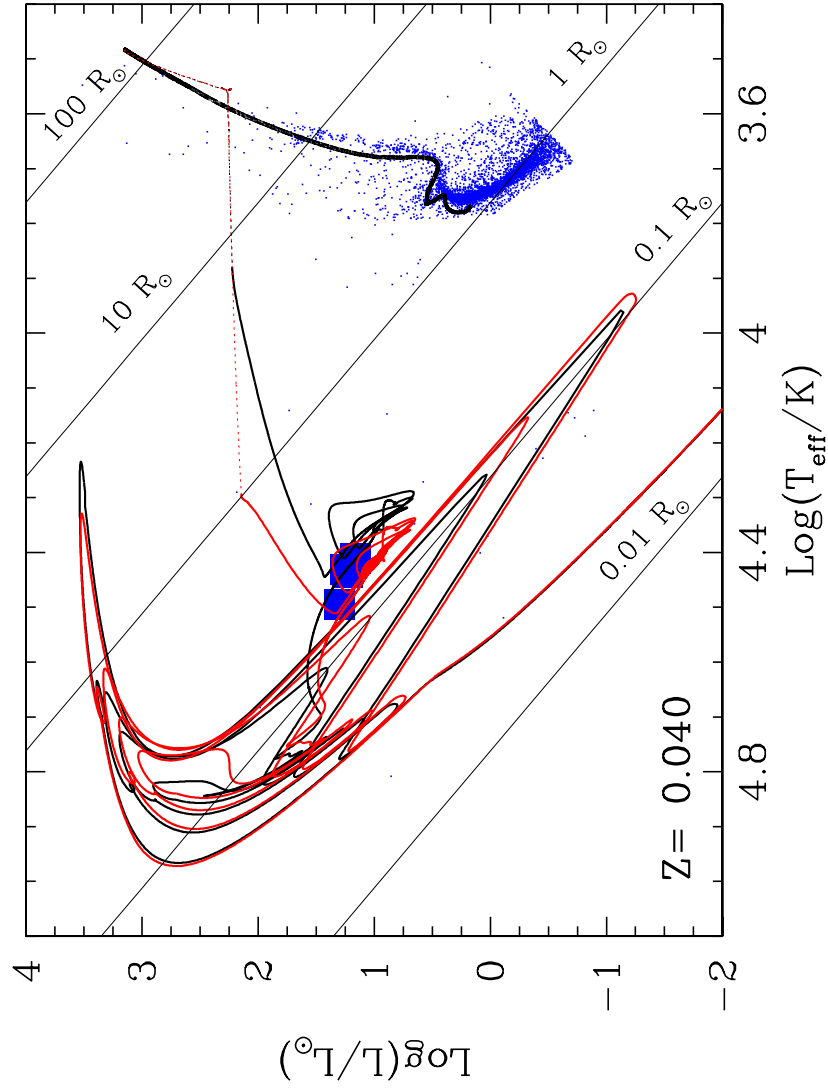}
\caption{Evolutionary tracks for initial 1.3~$M_{\odot}$ model up to the moment at which it has a helium core with 0.4067~$M_{\odot}$. Since then it is assumed that the object undergoes a CE episode. If the star emerges from the CE with a radius of 7.5~(1)~$R_{\odot}$, it has a mass of 0.41344~(0.40973)~$M_{\odot}$ and follows the tracks depicted with black (red) solid line. The blue points are {\it bona fide} NGC 6791 members (see Buzzoni et al. 2012) and blue squares represent the EHB stars. \label{fig:HR}} \end{center}
\end{figure}

The evolutionary tracks of the two most massive models for each $R_{d}$ value are presented in Figure~\ref{fig:HR}. As noted above, two radii were assumed after the emergence from the CE phase. The larger $R_{d}$ value implies a thicker hydrogen rich layer,  and thus a lower T$_{\rm{eff}}$ during most of  the evolution.  At post-CE stages, the star evolves blue-ward and ignites helium off-center owing to strong neutrino emission.  The evolutionary track subsequently bends downward almost at constant radius.  Thereafter the star depletes the helium core and then progressively the bottom of helium rich layers, following a cyclical-like trend.  The stars exhibit EHB conditions during that stage (notice the blue squares in Figure~\ref{fig:HR} that represent the EHB stars in NGC~6791. When helium burning becomes weaker, the star contracts, again evolving blue-ward and igniting the outermost hydrogen layers that gives rise to few thermonuclear flashes. These flashes burn enough hydrogen to cause the star to finally evolve to the white dwarf stage. Lower mass objects undergo a larger number of cycles during helium burning and hydrogen flashes because nuclear ignition episodes are weaker.

\begin{figure} \begin{center}
\includegraphics[scale= 0.60,angle=0]{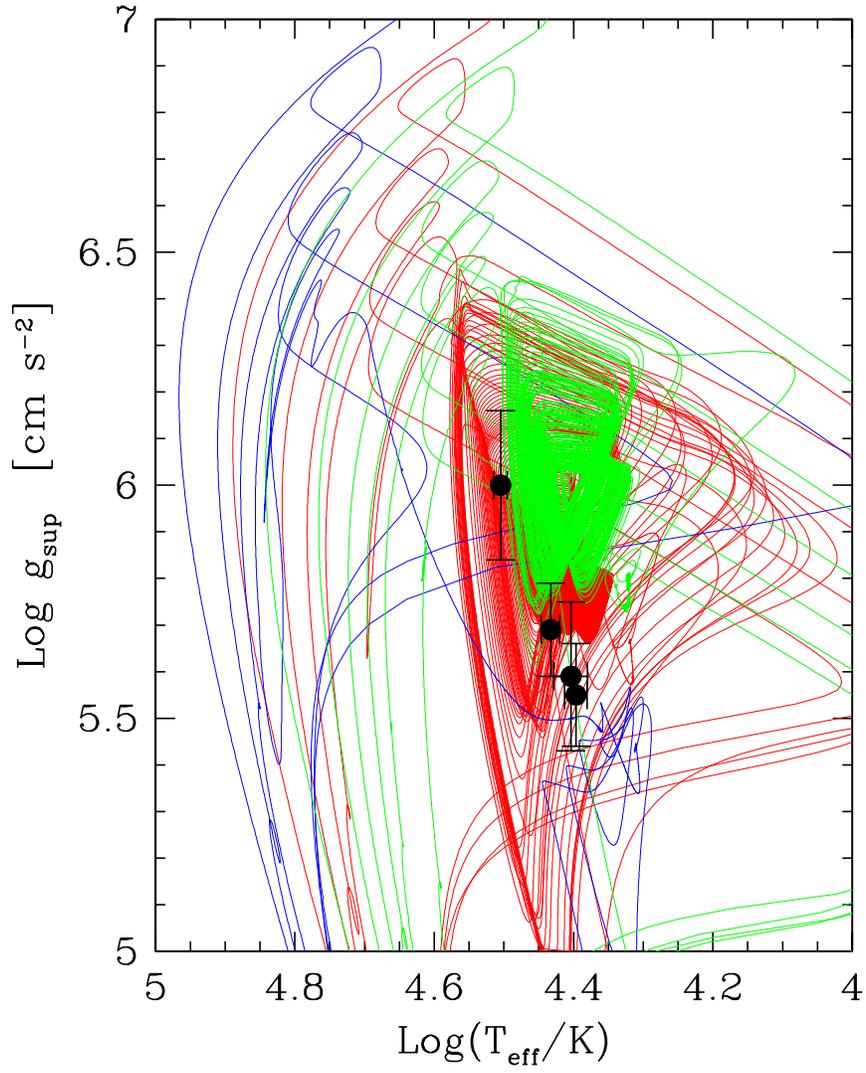}
\caption{Surface gravity as a function of T$_{\rm{eff}}$ for models that detach from CE with 7.5~$R_{\odot}$. Green, red and solid lines correspond to models with masses of 0.35048, 0.37367, and 0.41344~$M_{\odot}$, respectively. Data corresponding to stars B3-B6 is shown with their corresponding error bars. \label{fig:GTeff75}} \end{center}
\end{figure}

\begin{figure} \begin{center}
\includegraphics[scale= 0.60,angle=0]{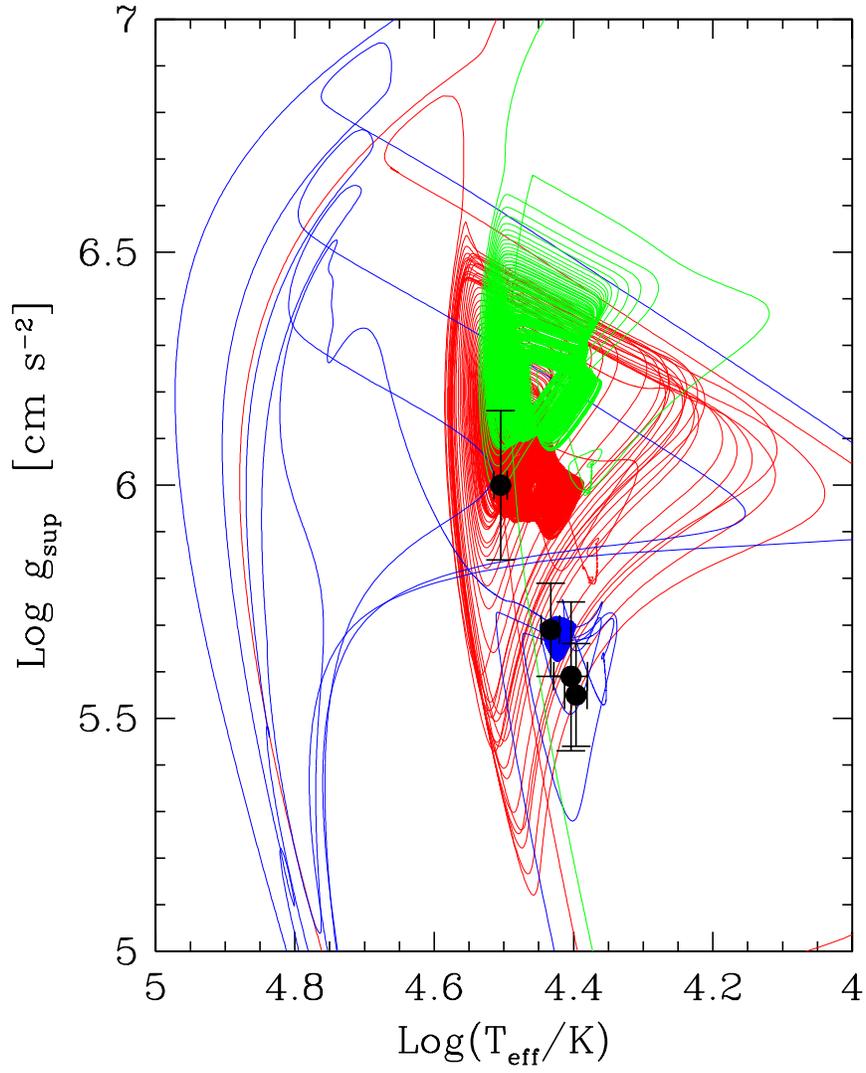}
\caption{Same as Figure~\ref{fig:GTeff75} but for the case in which CE resumes at 1~$R_{\odot}$. Here  green, red, and blue solid lines correspond to models with masses of 0.34863, 0.37084, and 0.40973~$M_{\odot}$, respectively. \label{fig:GTeff1}} \end{center}
\end{figure}

The evolution of stars that emerge from the CE phase with $R_{d}= 7.5\; R_{\odot}$ is shown in Figure~\ref{fig:GTeff75}, together with data corresponding to the EHB stars B4-B7.  Successfully, the model produces T$_{\rm{eff}}$ and surface gravities that match the observations. This is largely due to helium ignition that makes the star to stop its contraction at the right conditions.As expected, stars that emerge from the CE stage featuring $R_{d}= 1\; R_{\odot}$ exhibit a larger surface gravity since they are more compact (Figure~\ref{fig:GTeff1}). 

\begin{figure} \begin{center}
\includegraphics[scale= 0.60,angle=0]{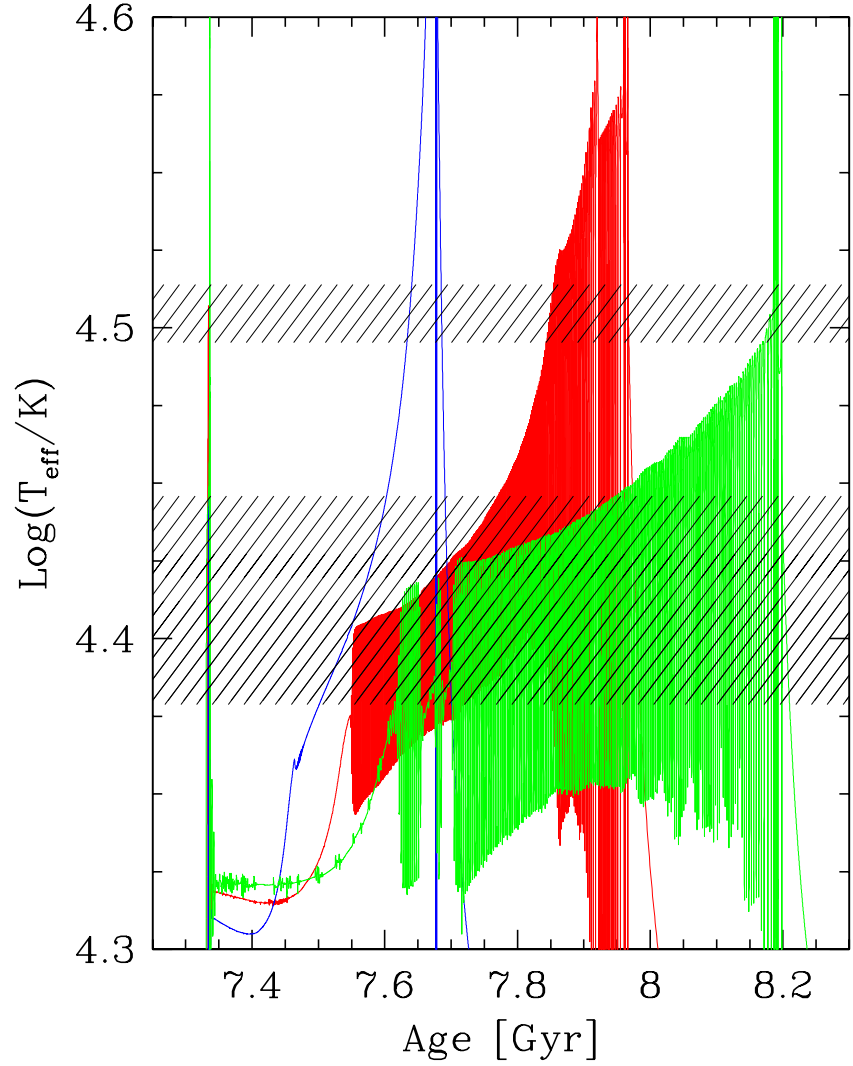}
\caption{Effective temperature as a function of time for the models included in Figure~\ref{fig:GTeff75}. As there, green, red, and blue solid lines correspond to models with masses of 0.35048, 0.37367, and 0.41344~$M_{\odot}$, respectively. {\bf The hatched areas indicate the T$_{\rm{eff}}$ interval (due to error bars) as given in} \citet{1994AJ....107.1408L}. B3, B4, and B5 intervals overlap in the lower hatched region meanwhile the upper one corresponds to the star B6. Notice that the T$_{\rm{eff}}$ of the models fall inside the observed intervals during a considerable amount of time, making its detection probable. \label{fig:tTeff75}} \end{center}
\end{figure}

\begin{figure} \begin{center}
\includegraphics[scale= 0.60,angle=0]{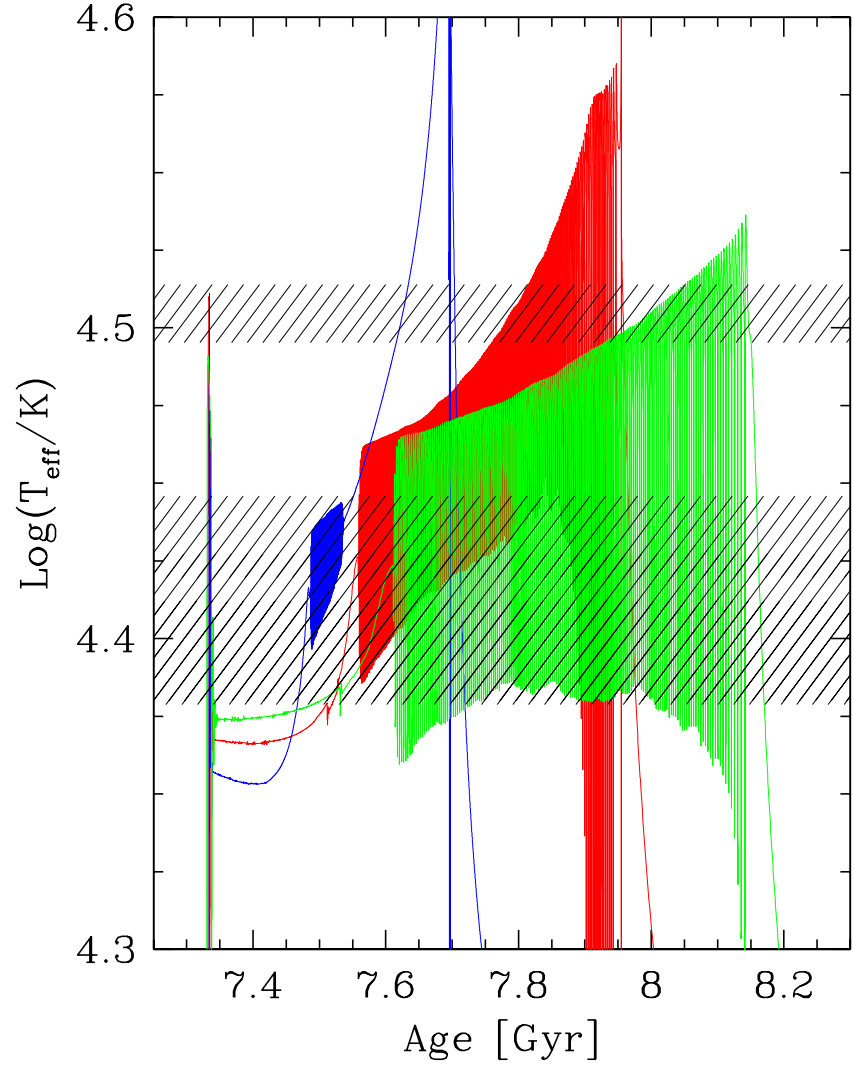}
\caption{Same as Figure~\ref{fig:tTeff75} but for the case in which CE resumes at 1~$R_{\odot}$. \label{fig:tTeff1}} \end{center}
\end{figure}

It can be noticed from Figure~\ref{fig:HR} that tracks pass several times across the T$_{\rm{eff}}$ interval corresponding to EHBs ($(\Delta T_{\rm{eff}})_{\rm{EHB}}$); see Section~\ref{sec:intro}.  Most of the time they fall at $(\Delta T_{\rm{eff}})_{\rm{EHB}}$ they undergo helium burning dominated cycles. The time they spend at $(\Delta T_{\rm{eff}})_{\rm{EHB}}$ during thermonuclear flashes and the final white dwarf cooling track is much shorter. So, the time the modelled stars spend at $(\Delta T_{\rm{eff}})_{\rm{EHB}}$ is essentially that when they resemble EHBs. This time  is crucial since the longer the time the easier to find them as EHBs.
Figure~\ref{fig:tTeff75} conveys the temporal evolution as a function of T$_{\rm{eff}}$ for the case of $R_{d}= 7.5\; R_{\odot}$. Temperature intervals indicated by the observations presented in \citet{li94} are likewise included. Remarkably, the modelled stars can be detected as EHBs for several hundred million years. The same is true for models featuring $R_{d}= 1\; R_{\odot}$ (see Figure~\ref{fig:tTeff1}). 

%------------------------------------------------------------------
\section{Discussion} \label{sec:disc}

The resulting orbital periods of such systems can be estimated via Equation~3 in \citet{iv13}:
\begin{equation}
 \frac{G M_1 M_{1,env}}{\lambda R_1}= \alpha_{CE} \bigg(
 - \frac{G M_1 M_2}{2 a_i} 
 + \frac{G M_{1,c} M_2}{2 a_f} 
 \bigg)
\end{equation}
where $G$ is the gravitational constant, $M_1$, $M_{1,env}$, $M_{1,c}$ are the total, envelope, and core masses of the donor star, respectively. $M_2$ is the companion mass, $a_i$, $a_f$ are the initial and final semi-axes, $\alpha_{CE}$ is the CE efficiency and $\lambda$ accounts for the density profile of the donor star.  The semi-axis at the onset of mass transfer is $a_i$, and is computed via the relation between the orbital semi-axis and the equivalent radius of the Roche lobe \citep{eg83}.  The final orbital period $P_f$ follows, and is an (increasing) function of the parameter $\xi= \lambda\; \alpha_{CE} / 2$. Here, $\xi= 0.10$ and $M_2= 1\; M_{\odot}$ are adopted as representative values and the models corresponding to the case of $R_d= 1\; R_{\odot}$.

If Roche lobe overflow occurs when the donor star develops a helium core of $0.3486\; M_{\odot}$ and exhibits a radius of 69~$R_{\odot}$, then $a_i= 172.1\; R_{\odot}$,  $a_f= 1.876\; R_{\odot}$ and $P_f=0.254$~days. If overflow occurs when the helium core is $0.4067\; M_{\odot}$ and features a radius of 137~$R_{\odot}$, then $a_i= 342\; R_{\odot}$,  $a_f= 4.61\; R_{\odot}$ and $P_f=0.962$~days. 

The estimated periods fall in the range of observations \citep[][and references therein]{gr01,ma04}, and indeed, EHB B4 displays an orbital period of $P=0.4$~days \citep{pa11}. 

%------------------------------------------------------------------
\section{Conclusions} \label{sec:concl}

In this study it is advocated that binary evolution is the source of the EHB population within NGC~6791,
in full similarity with field subdwarf \citep{ha02}. 
 An updated form of the \citet{bd03} code is used to demonstrate that EHBs can emerge from the post CE evolution of binary stars with masses conducive to the cluster's turn-off ($M_{to} \approx 1.15 \; M_{\odot}$).  The numerical model employed yields synthetic stars that match the observational and temporal properties of NGC~6791's EHB members.  The binary mechanism is not only means for stars to evolve to EHB conditions, since isolated stars with heavy mass loss might succeed. However, the evolutionary path explored here is preferred since it does not require ad-hoc anomalous and observationally unconfirmed mass-loss rates, and granted that NGC~6791 and EHB stars exhibit a high rate of binarity. 

One may wonder whether our results can be extended to other stellar systems. Unfortunately, no other open cluster is known to harbour EHB stars, which might be interpreted arguing that they are by far less massive than NGC 6791, even if they host a comparable amount of binary stars. This stresses once again the uniqueness of NGC 6791 among open clusters in the Milky Way. On the other hand, EHBs stars are more common in globular clusters, but they do no share the same properties of NGC 6791 EHB population \citep{li94} . First, in NGC 6791 EHBs are not centrally concentrated \citep{bu12}, while in globular they are \citep{li94}. Second, in globulars they span a much wider range in colours (hence temperature). 
This was historically  interpreted with the existence of a wide range of envelope sizes, hence  with differential mass loss during the RGB ascent. 
Nowadays the segmented EHB in globulars is mostly interpreted as an evidence of multiple stellar generations, 
each segment with different degree of He enhancement \citep{Ma14}.  Other authors consider rapid rotation \citep{Ta15}as well. These scenarios are difficult to invoke for NGC 6791 since we lack any accepted evidence of multiple stellar populations in NGC 6791 (see \citet{gei12} and \citet{bra14}).

%------------------------------------------------------------------
\acknowledgments

G.C. deeply thanks La Plata Observatory for financial support during a visit where this project was started. The authors deeply thank Daniel Majaess for reading and commenting on the manuscript.

\vspace{5mm}
%%\facilities{HST(STIS), Swift(XRT and UVOT), AAVSO, CTIO:1.3m, CTIO:1.5m,CXO}

%%\software{IRAF, cloudy, IDL}

%------------------------------------------------------------------

%%\allauthors

%%\listofchanges

%------------------------------------------------------------------

\begin{thebibliography}{}
\bibitem[Bedin et al.(2008)]{be08} Bedin, L.~R., Salaris, M., Piotto, G., Cassisi, S., Milone, A.~P., Anderson, J., King, I.~R. \ 2008, \apj, 679, L29
\bibitem[Benvenuto \& De Vito(2003)]{bd03} Benvenuto, O.~G., \& De Vito, M.~A.\ 2003, \mnras, 342, 50 
\bibitem[Bragaglia et al.(2014)]{bra14} Bragaglia, A., Sneden, C., Carretta, E., Gratton, R., Lucatello, S., Bernath, P.F., James, S.A, Ram, R.S., 2014, \apj, 796, 68
\bibitem[Brown(2008)]{br08} Brown, D.\ 2008, Hot Subdwarf Stars and Related Objects, 392, 83 
\bibitem[Buzzoni et al.(2012)]{bu12}Buzzoni, A., Bertone, E., Carraro, G., Buson, L. \ 2012, \apj, 749, 35
\bibitem[Carraro \& Chiosi(1995)]{cc95} Carraro, G., Chiosi, C. \ 1995, IAA-IAC-University of Pisa Workshop: The formation of the Milky Way, 175
\bibitem[Carraro et al.(1996)]{ca96} Carraro, G., Girardi, L., Bressan, A., Chiosi, C. \ 1996, A\&A, 305, 849
\bibitem[D'Cruz et al.(1996)]{dc96} D'Cruz, N.~L., Dorman, B.,  Rood, R.~T., O'Connell, R.W., \ 1996, \apj, 466, 359
\bibitem[Eggleton(1983)]{eg83} Eggleton, P.~P.\ 1983, \apj, 268, 368 
\bibitem[Geisler et al. (2012)]{gei12}Geisler, D., Villanova, S., Carraro, G., Pilachowski, C., Cummings, I., Johanos, C.I., Bresolin, F., 2012, \apj, 756, L40
\bibitem[Green et al.(2001)]{gr01} Green, E.~M., Liebert, J., \& Saffer, R.~A.\ 2001, 12th European Workshop on White Dwarfs, 226, 192 
\bibitem[Han et al.(2002)]{ha02} Han, Z., Podsiadlowski, P., Maxted, P.~F.~L., Marsh, T.~R., \& Ivanova, N.\ 2002, \mnras, 336, 449 
\bibitem[Iben \& Tutukov (1993)]{it93} Iben, I., Jr., \& Tutukov, A.~V.\ 1993, \apj, 418, 343 
\bibitem[Ivanova et al.(2013)]{iv13} Ivanova, N., Justham, S., Chen, X., et al.\ 2013, \aapr, 21, 59 
\bibitem[Kalirai et al.(2007)]{ka07} Kalirai, J.Zs., Bergeron, P., Hansen, B.~M.~S., Kelson, D.~.D., Reitzel, D.~B., Rich, R.~M., Richer, H.~B., \ 2007, \apj, 671, 748 
\bibitem[Kaluzny \& Udalski(1992)]{ku92} Kaluzny, J., Udalski, A. \ 1992, Acta Astronomica, 42, 29
\bibitem[Kippenhahn et al.(1967)]{kwh67} Kippenhahn, R., Weigert, A., \& Hofmeister, E.\ 1967, Methods in Computational Physics, 7, 129 
\bibitem[Langer et al.(1985)]{la85} Langer, N., El Eid, M.~F., \& Fricke, K.~J.\ 1985, \aap, 145, 179 
\bibitem[Liebert et al.(1994)]{li94} Liebert, J., Safferm R.~A., Green, E.~M. \ 1994, \aj, 107, 1408
\bibitem[Marino et al. (2014)]{Ma14} Marino, A.F., Milone, A.P., Przybilla, N., Bergemann, M., Lind, K., Asplund, M., Cassisi, S., et al., 2014, \mnras, 437, 1609
\bibitem[Maxted(2004)]{ma04} Maxted, P.~F.~L.\ 2004, Spectroscopically and Spatially Resolving the Components of the Close Binary Stars, 318, 387 
\bibitem[Mengel et al.(1976)]{me76} Mengel, J.~G., Norris, J., \& Gross, P.~G.\ 1976, \apj, 204, 488
\bibitem[Michaud et al.(2007)]{2007ApJ...670.1178M} Michaud, G., Richer, J., \& Richard, O.\ 2007, \apj, 670, 1178 
\bibitem[Miglio et al.(2012)]{mi12} Miglio, A., Brogaard, K., Stello, D., Chaplin, W.~J., D'Antona, F., Montalban, J., Basu, S., et al., \ 2012, \mnras, 419, 2077
\bibitem[Mochejska et al.(2003)]{mo03} Mochejska B.~J., Stanek, K.-Z., Kaluzny, J.\ 2013, \aj, 125, 3175 
\bibitem[Moni Bidin et al.(2008)]{mbd08} Moni Bidin, C. Catelan, M., Altmann, M., 2008, \aap, 480, L1
\bibitem[Pablo et al.(2011)]{pa11} Pablo, H., Kawaler, S.~D., Green, E.~M.  \ 2011, \apj, 740, L47
\bibitem[Paczynski(1976)]{pa76} Paczynski, B.\ 1976, Structure and Evolution of Close Binary Systems, 73, 75 
\bibitem[Reed et al.(2012)]{2012MNRAS.427.1245R} Reed, M.~D., Baran, A., {\O}stensen, R.~H., Telting, J., \& O'Toole, S.~J.\ 2012, \mnras, 427, 1245 
\bibitem[Reimers(1975)]{re75} Reimers, D.\ 1975, Memoires of the Societe Royale des Sciences de Liege, 8, 369
\bibitem[Tailo et al. (2015)]{Ta15} Tailo, M., D'antona, F., Vesperini, E., di Crescienzo, M., Ventura, P., Milone, A.P., Bellini, A.,,
et al., 2015, Nature, 523, 318
\bibitem[Twarog et al.(2011)]{tw11} Twarog, B.~A., Carraro, G., Anthony-Twarog, B.~J.\  2011, \apj, 727, L7 
\bibitem[van den Berg et al.(2013)]{va13} van den Berg, M., Verbunt, F., Tagliaferri, G., et al.\ 2013, \apj, 770, 98 
\bibitem[van Loon et al.(2008)]{va08} van Loon, J.~Th., Boyer, M.~L., McDonald, I. \ 2008, \apj, 680, L49
\bibitem[Yong et al.(2000)]{yo00} Yong, H., Demarque, P., Yi, S. \ 2000, \apj, 539, 928
\end{thebibliography}
\end{document}